\documentclass{svjour3}                     
\smartqed  
\usepackage{graphicx}
\journalname{Few-Body Systems (FB20)}
\newcommand{\SU}{{\rm SU}}

\newcommand{\MeV}{{\rm MeV}}

\begin{document}

\title{
Charmed mesons in nuclei with heavy-quark spin symmetry 
\thanks{Presented at the 20th International IUPAP Conference on Few-Body Problems in Physics, 20 - 25 August, 2012, Fukuoka, Japan}
}


\author{
L.Tolos \and C.Garcia-Recio \and J.Nieves \and O.Romanets \and L.L.Salcedo}


\institute{ L.Tolos \at
              Instituto de Ciencias del Espacio (IEEC/CSIC), Campus Universitat 
Aut\`onoma de Barcelona, Facultat de Ci\`encies, Torre C5, E-08193 Bellaterra 
(Barcelona), Spain\\
Frankfurt Institute for Advanced Studies, Johann Wolfgang Goethe University, Ruth-Moufang-Str. 1,
60438 Frankfurt am Main, Germany \\
              \email{tolos@ice.csic.es}           
           \and
            C. Garcia-Recio \at
           Departamento de F{\'\i}sica At\'omica, Molecular y Nuclear, 
Universidad de Granada, E-18071 Granada, Spain         \\
             \and
             J. Nieves \at
             Instituto de F{\'\i}sica Corpuscular (centro mixto CSIC-UV),
Institutos de Investigaci\'on de Paterna, Aptdo. 22085, 46071, Valencia, Spain \\
             \and
             O. Romanets \at
             Theory Group, KVI, University of Groningen,
Zernikelaan 25, 9747 AA Groningen, The Netherlands\\
             \and             
             L.L. Salcedo \at 
             Departamento de F{\'\i}sica At\'omica, Molecular y Nuclear, 
Universidad de Granada, E-18071 Granada, Spain         \\
}

\date{Received: date / Accepted: date}

\maketitle

\begin{abstract}
We study the properties of charmed pseudoscalar and vector mesons in dense matter within a unitary meson-baryon coupled-channel model which incorporates heavy-quark  spin symmetry. This is accomplished by extending the SU(3) Weinberg-Tomozawa Lagrangian to SU(8) spin-flavor symmetry and implementing a suitable flavor symmetry breaking. Several resonances  with negative parity are generated dynamically by the $s$-wave interaction between pseudoscalar and vector meson multiplets with $1/2^+$ and $3/2^+$ baryons. Those states are then compared to experimental data as well as theoretical models. Next, Pauli-blocking effects and meson self-energies  are introduced in a self-consistent manner to obtain the open-charm meson spectral functions in a dense nuclear environment. We finally discuss  the formation of $D$-mesic nuclei.
\end{abstract}

\section{Introduction}

The interest on the properties of open and hidden charmed mesons was triggered in the context of relativistic nucleus-nucleus collisions in connection to the charmonium suppression \cite{matsui} as a probe for the formation of Quark-Gluon Plasma (QGP).
The primary theoretical effort is to understand the interaction between hadrons with the charm degree of freedom. The nature of new charmed and strange hadron resonances is an active topic of research, with recent data coming from  CLEO, Belle, BaBar  and other experiments \cite{facility00}. In the upcoming years, the experimental program of the future FAIR facility at GSI \cite{gsi00} will move from the light quark sector to the heavy one and will face new challenges where charm plays a dominant role. 

Approaches based on coupled-channel dynamics have proven to be very successful in describing the existing experimental data. In particular, unitarized coupled-channel methods have been applied in the meson-baryon sector with charm content \cite{Tolos:2004yg,Lutz:2003jw,Hofmann:2005sw,Hofmann:2006qx,Mizutani:2006vq,JimenezTejero:2009vq}, partially motivated by the parallelism between the $\Lambda(1405)$ and the $\Lambda_c(2595)$. Other existing coupled-channel approaches are based on the J\"ulich meson-exchange model \cite{Haidenbauer:2007jq,Haidenbauer:2010ch} or on the hidden gauge formalism \cite{Wu:2010jy}.

However, these models are not fully consistent with heavy-quark spin symmetry (HQSS) ~\cite{Isgur:1989vq,Neubert:1993mb,Manohar:2000dt}, which is a proper QCD symmetry that appears when the quark masses, such as the charm mass, become larger than the typical confinement scale. Aiming at incorporating HQSS, an SU(8) spin-flavor symmetric model has been recently developed \cite{GarciaRecio:2008dp,Gamermann:2010zz}, similarly to the SU(6) approach in the light sector of Refs.~\cite{GarciaRecio:2005hy,Toki:2007ab}. The model can generate dynamically resonances with negative parity in all the isospin, spin, strange and charm sectors  that one can form from an s-wave interaction between pseudoscalar and vector meson multiplets with $1/2^+$ and $3/2^+$ baryons \cite{Romanets:2012hm}. 

In this work we will focus on the modifications of the dynamically-generated states in the nuclear medium and on the properties of open-charm mesons in dense nuclear matter. We will study the open-charm meson spectral functions in this nuclear environment within a self-consistent approach in coupled channels, and discuss their implications on the formation of $D$-mesic nuclei.

\section{SU(8) Weinberg-Tomozawa interaction with heavy-quark spin symmetry}
\label{su8wt}

In QCD all types of spin interactions vanish for infinitely massive quarks:  the dynamics is unchanged under arbitrary transformations in the spin of the heavy quark.  This is the prediction of HQSS. Thus, HQSS connects vector and pseudoscalar mesons containing charmed quarks.  Chiral symmetry fixes  the lowest order interaction between Goldstone bosons and other hadrons  in a model independent way; this is the Weinberg-Tomozawa (WT) interaction. Then, it is very appealing to have a predictive model for four flavors including all basic hadrons (pseudoscalar and vector mesons, and $1/2^+$ and $3/2^+$ baryons) which reduces to the WT interaction in the sector where Goldstone bosons are involved and which incorporates HQSS in the sector where charm quarks participate. Indeed, this is a model assumption which is justified in view of the reasonable semiqualitative outcome of the SU(6) extension in the three-flavor sector \cite{Gamermann:2011mq} and on a formal plausibleness on how the SU(4) WT interaction in the charmed pseudoscalar meson-baryon sector comes out in the vector-meson exchange picture.

The model obeys SU(8) spin-flavor symmetry and also HQSS in the sectors where the number of $c$- and $\bar{c}$- quarks are conserved separately. This is a model extension of the WT SU(3) chiral Lagrangian \cite{Romanets:2012hm}. Schematically,
\begin{equation}
\label{symbolic}
{\mathcal L}^{{\rm SU (8)}}_{ \rm WT} 
=
 \frac{1}{f^2} [[M^{\dagger} \otimes M]_{\bf 63_{a}}
  \otimes [B^{\dagger} \otimes B]_{ \bf 63} ]_{ \bf 1} ,
\end{equation}
which represents the interaction between baryons (in the 120 irrep of SU(8)) and mesons (in the 63) through $t$-channel exchange in the 63. In the $s$-channel, the meson-baryon space reduces into four SU(8) irreps, from which two multiplets ${\bf 120}$ and ${\bf 168}$ are the most attractive. As a consequence, dynamically-generated baryon resonances are most likely to occur in those sectors, and therefore we will concentrate on states which belong to these two representations. 

The SU(8)-extended WT meson-baryon interaction is given by
\begin{equation}
V_{ij}(s)= D_{ij}
\frac{2\sqrt{s}-M_i-M_j}{4\,f_i f_j} \sqrt{\frac{E_i+M_i}{2M_i}}
\sqrt{\frac{E_j+M_j}{2M_j}} 
\,.
\label{eq:vsu8break}
\end{equation}
Here, $i$ ($j$) are the outgoing (incoming) meson-baryon channels. The quantities $M_i$, $E_i$ and $f_i$ stand for the baryon mass and energy, in the center of mass frame, and the meson decay constant in the $i$ channel, respectively. $D_{ij}$  are the matrix elements coming from the SU(8) group structure of the coupling for the various charm, strange, isospin and spin ($CSIJ$) sectors, which display exact SU(8) invariance. However, this symmetry is severely broken in nature, so we implement a symmetry-breaking mechanism. The symmetry breaking pattern follows the chain $\SU(8)\supset\SU(6)\supset\SU(3)\supset\SU(2)$, where the last group refers to isospin. The symmetry breaking is introduced by means of a deformation of the mass and decay constant parameters.  This allows us to assign well-defined SU(8), SU(6) and SU(3) labels to the resonances and to find HQSS invariant states.

We solve the on-shell Bethe-Salpeter equation in coupled channels so as to calculate the scattering amplitudes, $T_{ij}$, by using the interaction matrix $V$ as kernel:
\begin{equation}
\label{LS}
 T(s) =\frac{1}{1-V(s)G(s)} V(s).
\end{equation}
$G(s)$ is a diagonal matrix containing the meson-baryon propagator for each channel.  $D$, $T$, $V$, and $G$ are matrices in coupled-channel space. The loop function $G(s)$ is logarithmically ultraviolet divergent and it is regularized by the means of a subtraction point prescription \cite{Nieves:2001wt}.

The poles of the scattering amplitudes are the dynamically-generated baryon resonances. Close to the pole, the $T$-matrix behaves as
\begin{equation}
\label{Tfit}
 T_{ij} (s) \approx \frac{g_i g_j}{\sqrt{s}-\sqrt{s_R}}
\,,
\end{equation}
with the mass ($m_R$) and the width ($\Gamma_R$) given by $\sqrt{s_R}=m_R - \rm{i} \Gamma_R/2$, while $g_i$ is the coupling to a given meson-baryon channel.

\section{Charmed and strange dynamically-generated baryon states}
\label{dyn}

Dynamically generated states in different charm and strange sectors  are predicted within our model \cite{GarciaRecio:2008dp,Gamermann:2010zz,Romanets:2012hm}. Some of them can be identified with known states from the PDG \cite{Nakamura:2010zzi}. This identification is made by comparing the PDG data on these states with the mass, width and, most important, the coupling to the meson-baryon channels of our dynamically-generated poles. Next we will concentrate on the $C=1,~S=0$ sector since the states generated in this sector are the most relevant ones for the latter analysis of open-charm mesons in dense nuclear matter.
 
We reproduce the three lowest-lying states of Ref.~\cite{GarciaRecio:2008dp} in the $\Lambda_c$ sector, which come from the most attractive SU(8) representations. However, those states have slightly different masses due to the different subtraction point, and the use of slightly different $D_s$ and $D_s^*$ meson decay  constants. The experimental $\Lambda_c(2595)$ resonance can be identified with the pole that we obtain around $2618.8\,\MeV$, as similarly done in Ref.~\cite{GarciaRecio:2008dp}. The width in our case is, however, bigger because we have not changed the subtraction point to fit its position as done in Ref.~\cite{GarciaRecio:2008dp}.  Nonetheless, the dominant three-body decay channel $\Lambda_c \pi \pi$ \cite{Nakamura:2010zzi} is not included in our calculation. A second broad $\Lambda_c$ resonance at 2617~MeV is, moreover, observed with a large coupling to the open channel $\Sigma_c \pi$, very close to $\Lambda_c(2595)$. This is the same two-pole pattern found in the charmless $I = 0, S = -1$ sector for the $\Lambda(1405)$\cite{Jido:2003cb}. A third spin-$1/2$ $\Lambda_c$ resonance is seen around 2828~MeV  and cannot be assigned to any experimentally known resonance. With regard to spin-$3/2$ resonances, we find one located at $(2666.6 -i 26.7\,\MeV)$. As seen in Ref.~\cite{GarciaRecio:2008dp}, this resonance is assigned to the experimental $\Lambda_c(2625)$. The $t$-channel vector-exchange model of Ref.~\cite{Hofmann:2006qx} also observed a similar resonance  at $2660\,\MeV$. The novelty of our calculations with respect to that one is that we obtain a non-negligible contribution from the vector meson-baryon channels to the generation of this resonance.

For $C=1, S=0, I=1, J=1/2$ ($\Sigma_c$ sector), three $\Sigma_c$ resonances are obtained with masses 2571.5, 2622.7 and 2643.4~MeV and widths 0.8, 188.0 and 87.0~MeV, respectively. Those states nicely agree with the three lowest lying resonances found in Ref.~\cite{GarciaRecio:2008dp} and are pure predictions of our model. Ref.~\cite{JimenezTejero:2009vq} predicts the existence of two resonances in this sector. However, only one of them can be identified to one of ours but with a strong vector meson-baryon component. Moreover, we predict two spin-$3/2$ $\Sigma_c$ resonances. The first one, a bound state at 2568.4~MeV, is thought to be the charmed counterpart of the $\Sigma(1670)$. The second state at $2692.9 -i 33.5$ ~MeV has not a direct experimental comparison.

\section{Open-charm mesons in nuclear matter}
\label{medium}

We start this section by first analyzing the behavior of the dynamically-generated states within the $C=1$ and $S=0$ sector in dense nuclear matter. Then, we study the properties of open charm ($D$ and $D^*$) mesons in a dense nuclear environment related to the modification of dynamically-generated  $\Lambda_c$ and $\Sigma_c$ states in this environment.

The self-energy and, hence, spectral function for $D$ and $D^*$ mesons are obtained self-consistently in a simultaneous manner, as it follows from HQSS by taking, as bare interaction, the SU(8)-extended WT interaction described in Sec.~\ref{su8wt}. We incorporate Pauli blocking effects and open charm meson self-energies in the intermediate propagators \cite{tolos09} for the in-medium solution.

\begin{figure}
\begin{center}
\includegraphics[width=0.43\textwidth,height=5.5cm]{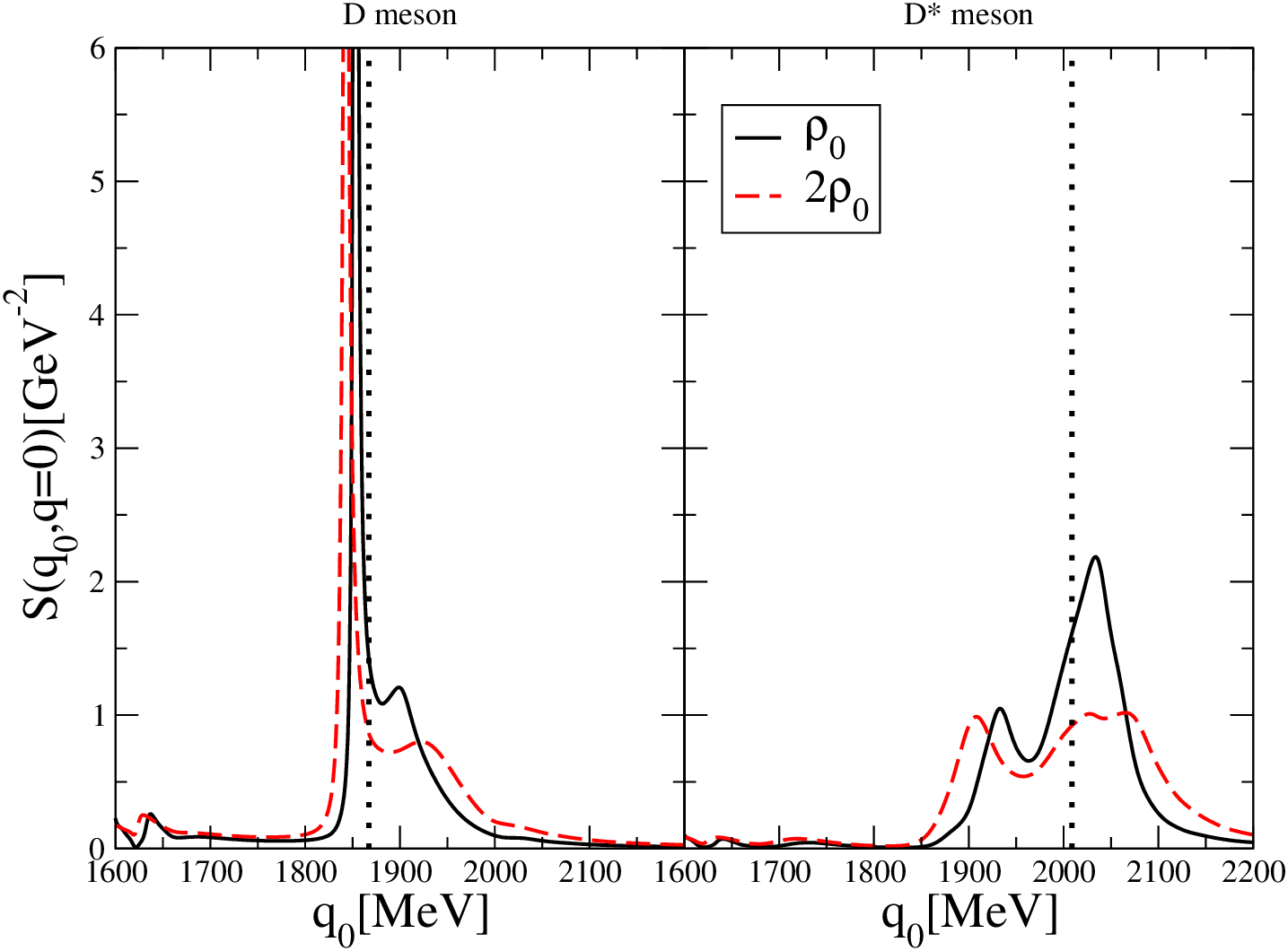}
\hfill
\includegraphics[width=0.43\textwidth, angle=90,height=5.3cm]{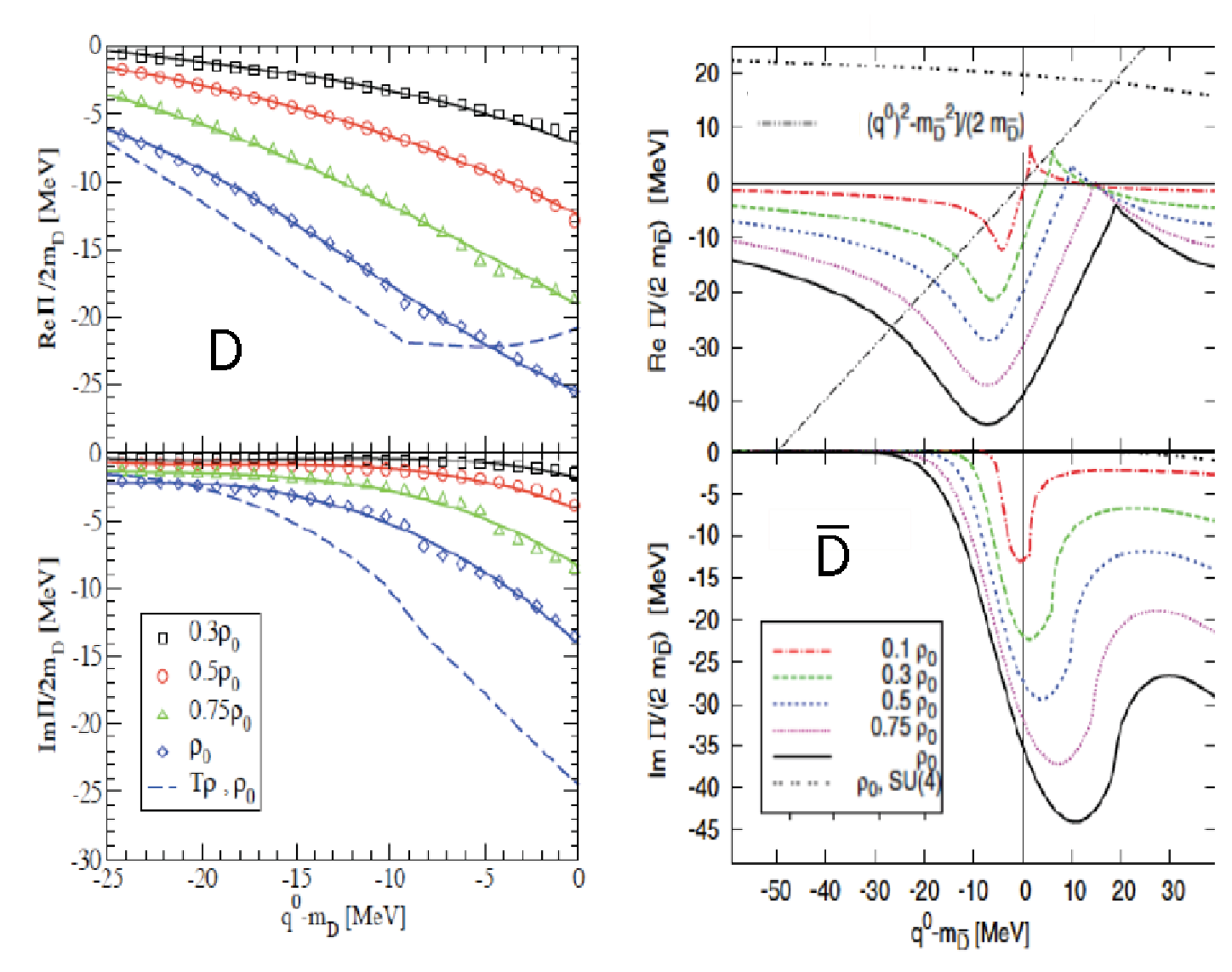}
\caption{Left: The $D$ and $D^*$ spectral functions in dense nuclear matter at $\vec{q}=0$ MeV/c. Right:  The $D$ and $\bar D$ optical potential at $\vec{q}=0$ MeV/c for different densities }
\label{fig1}
\end{center}
\end{figure}


The $D$ and $D^*$ self-energies are obtained by summing the transition amplitude for the different isospins over the Fermi sea of nucleons, $n(\vec{p})$: 
\begin{eqnarray}
\Pi_D(q_0,{\vec q}) &=& \int \frac{d^3p}{(2\pi)^3}\, n(\vec{p})  \, [\, {T_{DN}}^{(I=0,J=1/2)} +
3 \, {T_{DN}}^{(I=1,J=1/2)}\, ]  \ ,  \nonumber \\ 
\Pi_{D^*}(q_0,\vec{q}\,) &=& \int \frac{d^3p}{(2\pi)^3} \, n(\vec{p}\,) \,
\, \left [~ \frac{1}{3} \, {T}^{(I=0,J=1/2)}_{D^*N}+
{T}^{(I=1,J=1/2)}_{D^*N}+ \right . \nonumber \\
&&  \left . \frac{2}{3} \,
{T}^{(I=0,J=3/2)}_{D^*N}+ 2 \,
{T}^{(I=1,J=3/2)}_{D^*N}\right ] \  ,
\label{eq:selfd}
\end{eqnarray}
\noindent
where the $T$-matrices are evaluated at $s=P_0^2-{\vec P}^2$, being  $P_0=q_0+E_N(\vec{p})$ and $\vec{P}=\vec{q}+\vec{p}$ are the total energy and momentum of the meson-nucleon pair in the nuclear matter rest frame, and ($q_0$,$\vec{q}\,$) and ($E_N$,$\vec{p}$\,) stand  for the energy and momentum of the meson and nucleon, respectively, in this frame. The self-energy is determined self-consistently since it is obtained from the
in-medium amplitude which contains the meson-baryon loop function, and this quantity itself is a function of the self-energy. Then, the meson spectral functions for $D$ and $D^*$ mesons read
\begin{eqnarray}
S_{D(D^*)}(q_0,{\vec q})= -\frac{1}{\pi}\frac{{\rm Im}\, \Pi_{D(D^*)}(q_0,\vec{q})}{\mid
q_0^2-\vec{q}\,^2-m_{D(D^*)}^2- \Pi_{D(D^*)}(q_0,\vec{q}) \mid^2 } \ .
\label{eq:spec}
\end{eqnarray}

\begin{figure}
\begin{center}
\includegraphics[width=0.34\textwidth,angle=-90]{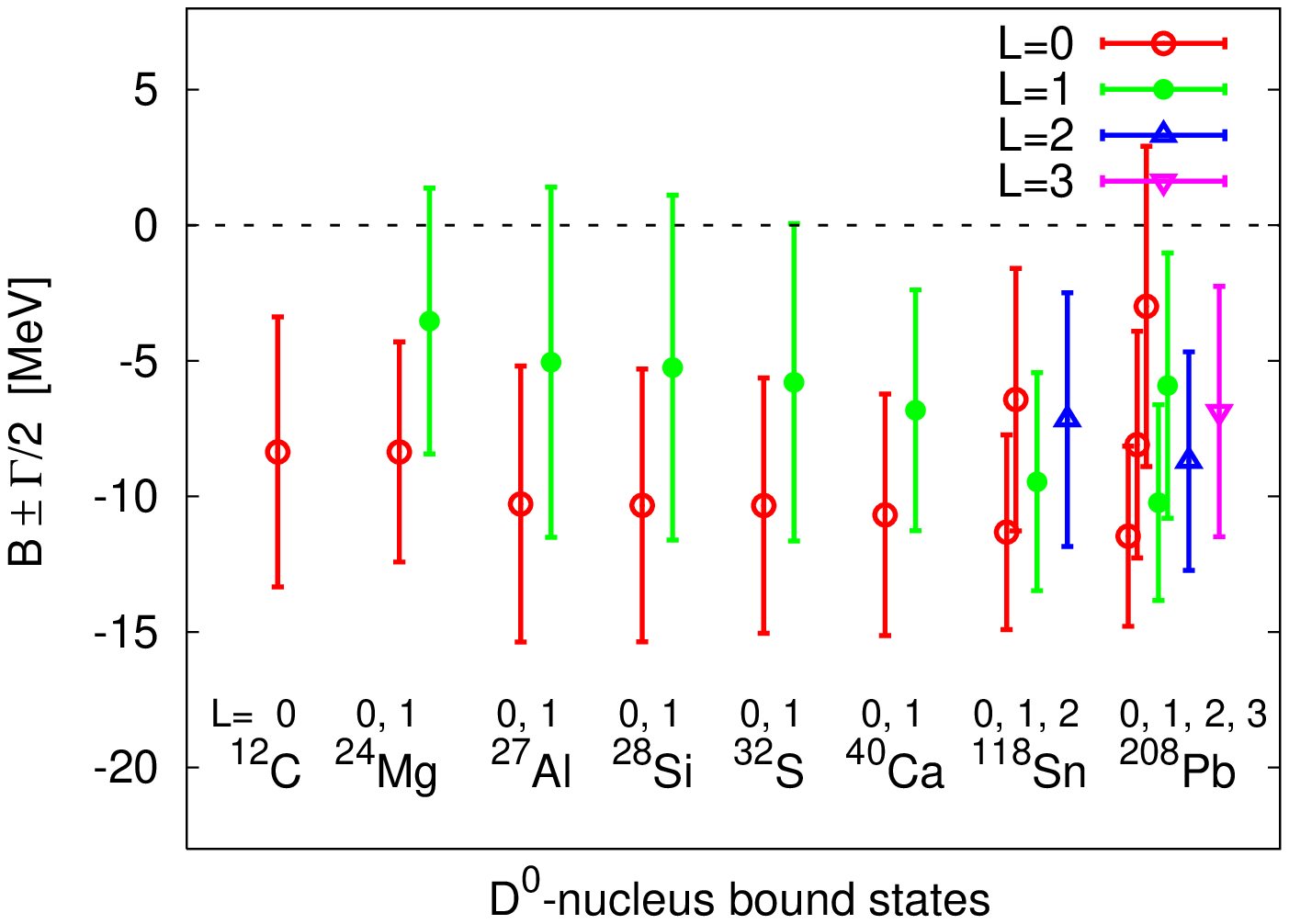}
\caption{$D^0$-nucleus bound states for different angular momentum $L$. \label{fig2}}
\end{center}
\end{figure}
The $D$ and $D^*$ spectral functions are displayed on the l.h.s. of  Fig.~\ref{fig1}. Those spectral functions show a rich spectrum of resonance-hole states. On one hand, the $D$ meson quasiparticle peak mixes strongly with $\Sigma_c(2823)N^{-1}$
and $\Sigma_c(2868)N^{-1}$ states. On the other hand, the $\Lambda_c(2595)N^{-1}$ is clearly visible in the low-energy tail.  With regard to the $D^*$ meson, the $D^*$ spectral function incorporates the $J=3/2$ resonances, and the quasiparticle peak fully mixes with the  $\Sigma_c(2902)N^{-1}$ and $\Lambda_c(2941)N^{-1}$ states.  For both mesons, the $Y_cN^{-1}$ modes tend to smear out and the spectral functions broaden with increasing phase space, as seen before in the $SU(4)$ model \cite{Mizutani:2006vq}. Note, however, that not all the states described in Sec.~\ref{dyn} are seen in the $D$ and $D^*$ spectral functions. This is due to the fact that some of those states do not strongly couple to $DN$ and/or $D^*N$ states. Moreover, as we have just seen, resonances with higher masses than those described in Sec.~\ref{dyn} are also present in the spectral functions. Those resonant states were seen in the wider energy range explored in Ref.~\cite{GarciaRecio:2008dp}.

\begin{figure}
\begin{center}
\includegraphics[width=0.47\textwidth]{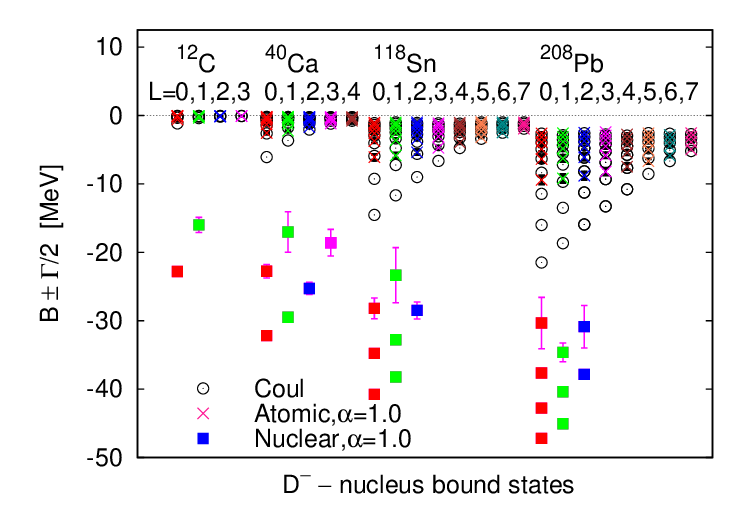}
\hfill
\includegraphics[width=0.47\textwidth]{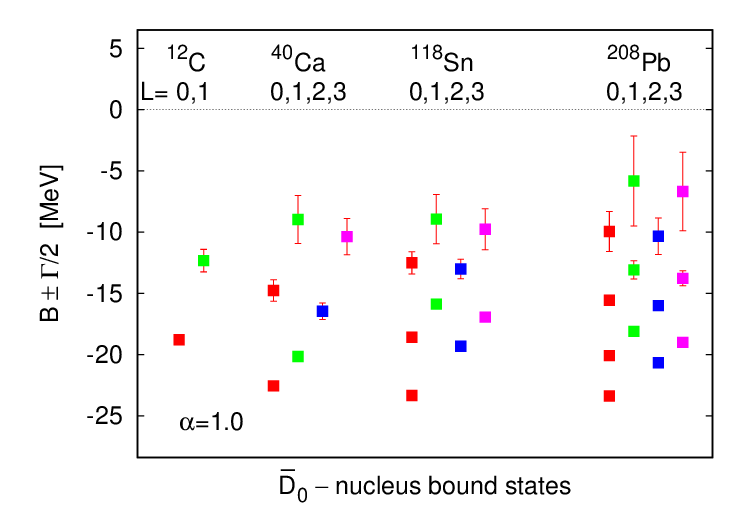}
\caption{$D^-$ and $\bar D^0$- nucleus bound states for different angular momentum $L$. \label{fig3}}
\end{center}
\end{figure}


\section{Charmed mesons in nuclei}

$D$ and $\bar D$-meson bound states in $^{208}$Pb were predicted in Ref.~\cite{tsushima99} relying upon an attractive  $D$ and $\bar D$ -meson potential in the nuclear medium. This potential was obtained within a quark-meson coupling (QMC) model \cite{sibirtsev99}. The experimental observation of those bound states, though, might be problematic since, even if there are bound states, their widths could be very large compared to the separation of the levels. This is indeed the case for the potential derived from a $SU(4)$ $t$-vector meson exchange model for $D$-mesons \cite{TOL07}.

We solve the Schr\"odinger equation in the local density approximation so as to analyze the formation of bound states with charmed mesons in nucleus. We use the energy dependent optical potential 
\begin{equation}
  V(r,E) = \frac{
  \Pi(q^0=m+E,\vec{q}=0,~\rho(r))}{2 m},
\label{eq:UdepE}
\end{equation}
where $E=q^0-m$ is the $D$ or $\bar D$ energy excluding its mass, and $\Pi$ the meson self-energy. The optical potential for different densities is displayed on the r.h.s of Fig.~\ref{fig1}. For $D$ mesons we observe a strong energy dependence of the potential close to the $D$ meson mass due to the mixing of the quasiparticle peak with the  $\Sigma_c(2823)N^{-1}$ and $\Sigma_c(2868)N^{-1}$ states. As for the $\bar D$ meson, the presence of a bound state at 2805 MeV \cite{Gamermann:2010zz}, almost at $\bar D N$ threshold, makes the potential also strongly energy dependent. This is in sharp contrast to the SU(4) model (see analysis in Ref.~\cite{carmen10}).

Then, the question is whether $D$ and/or $\bar D$  will be bound in nuclei. We start by discussing  $D$ mesons in nuclei.  We observe that the $D^0$-nucleus states are weakly bound (see Fig.~\ref{fig2}), in contrast to previous results using the QMC model. Moreover,  those states have significant widths \cite{carmen10}, in particular, for $^{208}$Pb \cite{tsushima99}. Only $D^0$-nucleus bound states are possible since the Coulomb interaction prevents the formation of observable bound states for $D^+$ mesons.

Apropos of  $\bar D$-mesic nuclei, not only $D^-$ but also $\bar{D}^0$ bind in nuclei as seen in Fig.~\ref{fig3}. The spectrum 
contains states of atomic and of nuclear types for all nuclei for $D^-$  while, as expected, only nuclear states are present for $\bar{D}^0$ in nuclei. Compared to the pure Coulomb levels, the atomic states are less bound. The nuclear ones are more bound and may present a sizable width \cite{GarciaRecio:2011xt}. Moreover, nuclear states only exist for low angular momenta.



The information on bound states is very valuable to gain some knowledge on the charmed meson-nucleus interaction, which is of interest for PANDA at FAIR. The experimental detection of $D$ and $\bar D$-meson bound states is, though, a difficult task. For example, reactions with antiprotons on nuclei for obtaining $D^0$-nucleus states 
might have a very low production rate (see Ref.~\cite{carmen10} for details). Reactions but with proton beams, although difficult, seem more likely to trap a $D^0$ in nuclei \cite{carmen10}.

\section{Conclusions and Future}

The properties of charmed mesons in infinite dense matter and nuclei have been studied within a unitary meson-baryon coupled-channel model which incorporates heavy-quark spin symmetry. Several resonances in the $C=1$ and $S=0$ sector have been analyzed and compared with experimental data from several facilities as well as with other theoretical models. The in-medium modifications on those baryonic resonances and on charmed mesons have been obtained. The in-medium solution  accounts for Pauli blocking effects and meson self-energies. We have analyzed the evolution with density of the open-charm meson spectral functions and finally studied the possible formation of $D$-mesic nuclei. On one hand, only  weakly bound $D^0$-nucleus states seem to be feasible. On the other hand, $D^-$ and $\bar D^0$- nuclear bound states are possible, the latter ones with a stronger binding than for $D^0$ and with also a sizable width. The challenge is, though, the experimental detection.

\begin{acknowledgements}
This research was supported by DGI and FEDER
funds, under Contract Nos. FIS2011-28853-C02-02,
 FIS2011-24149, FPA2010-16963 and the
Spanish Consolider-Ingenio 2010 Programme CPAN
(CSD2007-00042), by Junta de Andalucõa Grant
No. FQM-225, by Generalitat Valenciana under Contract
No. PROMETEO/2009/0090 and by the EU
HadronPhysics2 project, Grant Agreement No. 227431.
O. R.  wishes to acknowledge support from the
Rosalind Franklin Programme. L. T. acknowledges support
from Ramon y Cajal Research Programme, and from FP7-
PEOPLE-2011-CIG under Contract No. PCIG09-GA-
2011-291679.
\end{acknowledgements}



%
%







\end{document}